\documentclass[aps,prd,twocolumn,eqsecnum,showpacs,amsmath]{revtex4}
\usepackage[dvips]{color,graphicx}
\usepackage{amsfonts}
\usepackage{amssymb}

\newcommand{\dalm}{\kern1pt\vbox{\hrule height 0.9pt\hbox{\vrule width
0.9pt\hskip 2.5pt\vbox{\vskip 5.5pt}\hskip 3pt\vrule width 0.3pt}\hrule height
0.3pt}\kern1pt}
\newcommand{\ma}[1]{\mbox{$\mathcal{#1}$}}

\begin{document}


\title{
Effects of Gauss-Bonnet term on the final fate of gravitational collapse}

\author{
Hideki Maeda
\footnote{Electronic address: hideki@gravity.phys.waseda.ac.jp}
}

\address{ 
Advanced Research Institute for Science and Engineering,
Waseda University, Okubo 3-4-1, Shinjuku, Tokyo 169-8555, Japan
}

\date{\today}

\begin{abstract}                
We obtain a general spherically symmetric solution of a null dust fluid in $n (\geq 4)$-dimensions in Gauss-Bonnet gravity.
This solution is a generalization of the $n$-dimensional Vaidya-(anti)de Sitter solution in general relativity.
For $n=4$, the Gauss-Bonnet term in the action does not contribute to the field equations, so that the solution coincides with the Vaidya-(anti)de Sitter solution.
Using the solution for $n \ge 5$ with a specific form of the mass function, we present a model for a gravitational collapse in which a null dust fluid radially injects into an initially flat and empty region. 
It is found that a naked singularity is inevitably formed and its properties are quite different between $n=5$ and $n \ge 6$.
In the $n \ge 6$ case, a massless ingoing null naked singularity is formed, while in the $n=5$ case, a massive timelike naked singularity is formed, which does not appear in the general relativistic case.
The strength of the naked singularities is weaker than that in the general relativistic case.
These naked singularities can be globally naked when the null dust fluid is turned off after a finite time and the field settles into the empty asymptotically flat spacetime.
\end{abstract}


\maketitle

\section{Introduction}
The final fate of gravitational collapse is one of the most important unsolved problems in gravitation physics.
In general relativity, singularity theorems were proved stating that spacetime singularities inevitably appear under general situations and physical energy conditions~\cite{he1973}. 
In this context, a {\em cosmic censorship hypothesis} (CCH) was proposed by Penrose, which asserts that singularities formed in generic gravitational collapse of physical matter cannot be observed; in other words, there are no {\em naked singularities} formed in physical gravitational collapse~\cite{penrose1969,penrose1979}.
There are two versions of this hypothesis. 
The weak CCH asserts that all singularities formed in gravitational collapse are hidden inside black holes and cannot be observed from infinity, which implies the future predictability of the spacetime outside the black-hole event horizons. 
On the other hand, the strong CCH asserts that no singularities are observable. 

At the present time, however, the general proof of CCH is far from complete; on the contrary, many counterexample candidates have been proposed in general relativity~\cite{harada2004}. 
The generic occurrence of naked singularities has been shown in the spherical dust collapse represented by the Lema\^{\i}tre-Tolman-Bondi (LTB) solution~\cite{LTB,christodoulou1984,newman1986,ns} and in the null dust collapse represented by the Vaidya solution~\cite{vaidya,JoshiDwivedi,JoshiDwivedi2,lake1991} although the absence of pressure is not physically reasonable.
For a perfect fluid with a reasonable equation of state $p=k\mu$, where $p$ and $\mu$ are the pressure and the energy density, respectively, the generic occurrence of naked singularities in spherical collapse has been shown by numerical simulations for $0<k\alt 0.0105$~\cite{harada1998,hm2001,op}.
Of course, the assumption of spherical symmetry is very strong.
In a genuine counterexample of CCH, the solution must be stable against non-spherical perturbations.
Stability analyses against non-spherical linear perturbations have been done for the marginally bound LTB solution~\cite{INH} and the self-similar Vaidya solution~\cite{nw2005}.
It has been shown numerically that the naked singularity formation in the LTB solution is marginally stable against odd-parity perturbations but is unstable against even-parity perturbations~\cite{INH}.
On the other hand, the naked singularity formation in the self-similar Vaidya solution has been shown to be stable against even-parity perturbations~\cite{nw2005}.
In the perfect-fluid case, stability of the naked singularity formation  against non-spherical perturbations is still open.

Singularity theorems imply that a stage with infinitely high curvature can be realized in the very final stage of gravitational collapse.
The effects of quantum gravity will be unavoidable there. 
Until now, many quantum theories of gravity have been proposed.
Among them, superstring theory is the most promising candidate, which predicts a higher-dimensional spacetime. 
If superstring theory is the correct quantum theory of gravity, the effects of the extra dimensions will become significant near singularity formation.
When the curvature radius around a center is comparable to the compactification radius of the extra dimensions, the collapse must be effectively higher-dimensional.

However, the non-perturbative aspects of superstring theory are so far not understood completely although the progress in recent years has been remarkable.
Given the present circumstances, taking string effects perturbatively into classical gravity is one possible approach to study the quantum effects of gravity.
The Gauss-Bonnet term in the Lagrangian is the higher curvature correction to general relativity and naturally arises as the next leading order of the $\alpha'$-expansion of heterotic superstring theory, where $\alpha'$ is the inverse string tension~\cite{Gross}.

Until now, the properties of static solutions in Gauss-Bonnet gravity have been intensively studied, particularly from the viewpoint of black-hole thermodynamics~\cite{GB_BH,Wheeler_1,GBstatic,massGB,tm2005}.
In comparison to this, gravitational collapse has been less investigated, and accordingly the effects of the Gauss-Bonnet term on the dynamical formation process of such a higher-dimensional black hole have not been clarified yet.
In this paper, considering the spherically symmetric gravitational collapse of a null dust fluid with the $n(\ge 4)$-dimensional action including the Gauss-Bonnet term for gravity, we investigate its effects on the final fate of gravitational collapse.

The organization of this paper is as follows.
In Section II, we introduce our model and obtain an exact solution which represents the gravitational collapse of a null dust fluid. 
In Section III, considering the situation in which a null dust fluid radially injects into an initially flat and empty region, we show that a naked singularity is inevitably formed.
In Section IV, the strength of the naked singularity is investigated. 
Section V is devoted to discussion and conclusions.
Throughout this paper we use units such that $c=1$. 
As for notation we follow \cite{Gravitation}. 
The Greek indices run $\mu=0,1, \cdots, n-1$.

\section{Model and solution}
We begin with the following $n$-dimensional ($n \geq 4$) action:
\begin{equation}
\label{action}
S=\int d^nx\sqrt{-g}\biggl[\frac{1}{2\kappa_n^2}(R-2\Lambda+\alpha{L}_{GB}) \biggr]+S_{\rm matter},
\end{equation}
where
$R$ and $\Lambda$ are the $n$-dimensional Ricci scalar and the cosmological constant, respectively. $\kappa_n$ is defined by $\kappa_n\equiv\sqrt{8\pi G_n}$, where $G_n$ is the $n$-dimensional gravitational constant.
The Gauss-Bonnet term ${L}_{GB}$ is the combination of the Ricci scalar, Ricci tensor $R_{\mu\nu}$, and Riemann tensor $R^\mu_{~~\nu\rho\sigma}$ as
\begin{equation}
{L}_{GB}=R^2-4R_{\mu\nu}R^{\mu\nu}+R_{\mu\nu\rho\sigma}R^{\mu\nu\rho\sigma}.
\end{equation}
In 4-dimensional spacetime, the Gauss-Bonnet term does not contribute to the field equations.
$\alpha$ is the coupling constant of the Gauss-Bonnet term. 
This type of action is derived in the low-energy limit of heterotic superstring theory~\cite{Gross}.
In that case, $\alpha$ is regarded as the inverse string tension and positive definite.
Therefore, only the case where $\alpha \ge 0$ is considered in this paper.
We consider a null dust fluid as a matter field, whose action is represented by $S_{\rm matter}$ in Eq.~(\ref{action}).

The gravitational equation of the action (\ref{action}) is
\begin{equation}
{G}^\mu_{~~\nu} +\alpha {H}^\mu_{~~\nu} +\Lambda \delta^\mu_{~~\nu}= \kappa_n^2 {T}^\mu_{~~\nu}, \label{beq}
\end{equation}
where 
\begin{eqnarray}
{G}_{\mu\nu}&=&R_{\mu\nu}-{1\over 2}g_{\mu\nu}R,\\
{H}_{\mu\nu}&=&2\Bigl[RR_{\mu\nu}-2R_{\mu\alpha}R^\alpha_{~\nu}-2R^{\alpha\beta}R_{\mu\alpha\nu\beta}
\nonumber
\\
&& ~~~~
 +R_{\mu}^{~\alpha\beta\gamma}R_{\nu\alpha\beta\gamma}\Bigr]
-{1\over 2}g_{\mu\nu}{L}_{GB}.
\end{eqnarray}
${H}_{\mu\nu} \equiv 0$ holds for $n=4$.
The energy-momentum tensor of a null dust fluid is
\begin{eqnarray}
{T}_{\mu\nu}=\rho l_{\mu}l_{\nu},
\end{eqnarray}
where $\rho$ is the non-zero energy density and $l_\mu$ is a null vector.

We consider the general spherically symmetric spacetime with the line element
\begin{equation}
\label{metric}
ds^2=-f(v,r)e^{-2\delta(v,r)}dv^2+2e^{-\delta}dvdr+S(v,r)^2d\Omega_{n-2}^2,
\end{equation}  
where $d\Omega_{n-2}^2$ is the line element of the $(n-2)$-dimensional unit sphere.
$v$ is the advanced time coordinate and hence a curve $v=\mbox{const.}$ denotes a radial ingoing null geodesic.
The normalization of $l_\mu$ is such that $l_{\mu}=-\partial_{\mu} v$.

If $S$ is constant, the $(r,v)$ component of the field equation (\ref{beq}) gives the contradiction $\rho=0$.
Thus, we hereafter adopt the gauge where $S=r$ without loss of generality.

The $(v,r)$ component of the field equation (\ref{beq}) gives $\delta'= 0$ or $f=1+r^2/[2\alpha(n-3)(n-4)]$, where the prime denotes the derivative with respect to $r$. 
In the latter case, which is valid only for $n \ge 5$, the $(r,v)$ component of Eq.~(\ref{beq}) gives a contradiction $\rho=0$, so that $\delta=\delta(v)$ is concluded.
We can set $\delta \equiv 0$ without loss of generality by rescaling the time coordinate.

Then, $f$ is obtained by solving only the $(v,v)$ or $(r,r)$ component of Eq.~(\ref{beq}), which is written as  
\begin{eqnarray}
&&rf'-(n-3)(1-f)+(n-1){\tilde \Lambda}r^2
\nonumber \\
&&~~~
+\frac{\tilde{\alpha}}{r^2}(1-f)\Bigl[2rf'-(n-5)(1-f)\Bigr]=0,
\end{eqnarray}  
where $\tilde{\alpha}\equiv (n-3)(n-4)\alpha$ and ${\tilde \Lambda}\equiv 2/[(n-1)(n-2)]\Lambda$.
This equation is integrated to give the general solution as
\begin{equation}
\label{f-eq}
f=1+\frac{r^2}{2\tilde{\alpha}}\Biggl\{1\mp\sqrt{1+4\tilde{\alpha}\biggl[\frac{m(v)}{r^{n-1}}+{\tilde \Lambda}\biggr]}\Biggr\} 
\end{equation}  
for $n\ge 5$, where $m(v)$ is an arbitrary function of $v$.
For $n=4$, the solution is 
\begin{equation}
f=1-\frac{m(v)}{r}-{\tilde \Lambda}r^2, \label{VdS}
\end{equation}  
which is the Vaidya-(anti)de Sitter solution.
In order for the term in the square root in Eq.~(\ref{f-eq}) to be non-negative,
\begin{equation}
m \ge -\left({\tilde \Lambda}+\frac{1}{4{\tilde \alpha}}\right)r^{n-1}
\end{equation}  
must be satisfied.
There are two families of solutions which correspond to the sign in front of the square root in Eq.~(\ref{f-eq}).
We call the family which has the minus (plus) sign the minus-branch (plus-branch) solution.
From the $(r,v)$ component of Eq.~(\ref{beq}), we obtain the energy density of the null dust fluid as 
\begin{eqnarray}
\rho=\frac{n-2}{2\kappa_n^2r^{n-2}}{\dot m} \label{density}
\end{eqnarray}  
for both branches, where the dot denotes the derivative with respect to $v$.
In order for the energy density to be non-negative, ${\dot m} \ge 0$ must be satisfied.

In the general relativistic limit ${\tilde \alpha} \to 0$, the minus-branch solution is reduced to 
\begin{eqnarray}
f=1-\frac{m(v)}{r^{n-3}}-{\tilde \Lambda}r^2, \label{vaidyaGR}
\end{eqnarray}  
which is the $n$-dimensional Vaidya-(anti)de Sitter solution, while there is no such limit for the plus-branch solution.
In the static case ${\dot m}=0$, the solution (\ref{f-eq}) is reduced to the solution independently discovered by Boulware and Deser~\cite{GB_BH} and Wheeler~\cite{Wheeler_1}.

\section{Naked singularity formation}
\begin{figure}[tbp]
\includegraphics[width=.99\linewidth]{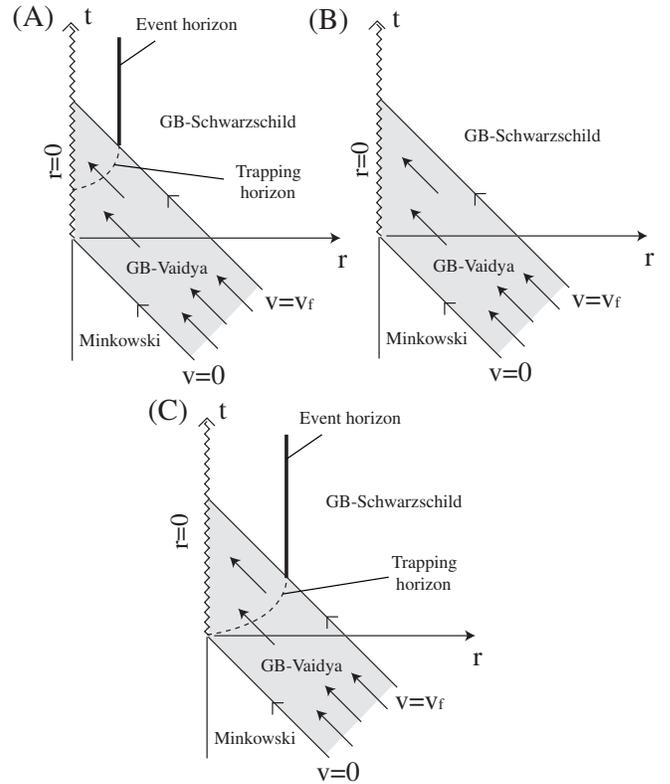}
\caption{
Gravitational collapse of a null dust fluid in (A) 5-dimension with an event horizon ($v_{\rm f} > v_{\rm AH}$), (B) 5-dimension without an event horizon ($v_{\rm f} \le v_{\rm AH}$), and (C) $n(\ge 6)$-dimension. 
A singularity is formed at $v=r=0$ and develops.
Here the Minkowski spacetime for $v<0$ is joined to the GB-Schwarzschild spacetime for $v>v_{\rm f}$ by way of the GB-Vaidya spacetime.
}
\label{Fig1}
\end{figure}

In this and the next sections, we study the gravitational collapse of a null dust fluid in Gauss-Bonnet gravity and compare it with that in general relativity by use of the solution obtained in the previous section.
We consider the minus-branch solution for $n \ge 5$ in order to compare with the general relativistic case.
For simplicity, we do not consider a cosmological constant, i.e., ${\tilde \Lambda}=0$.
In this case, Eq.~(\ref{f-eq}) is reduced to 
\begin{equation}
f=1+\frac{r^2}{2\tilde{\alpha}}\Biggl(1-\sqrt{1+4\tilde{\alpha}\frac{m(v)}{r^{n-1}}}\Biggr). \label{f-GB}
\end{equation}  
Hereafter we call the solution the Gauss-Bonnet-Vaidya (GB-Vaidya) solution. 
The special case in which $m$ is a constant we call the GB-Schwarzschild solution, of which the global structure is presented in~\cite{tm2005}.
We consider the situation in which a null dust fluid radially injects at $v=0$ into an initially Minkowski region ($m(v) = 0$ for $v<0$) (Fig.~\ref{Fig1}).
The form of $m(v)$ is assumed to be 
\begin{eqnarray}
m=m_0 v^{n-3}, \label{massform}
\end{eqnarray}  
where $m_0$ is a positive constant.
In this case, the solution~(\ref{f-GB}) is reduced to the $n$-dimensional self-similar Vaidya solution in the general relativistic limit ${\tilde \alpha} \to 0$~\cite{gs2000,gd2001}.

From Eq.~(\ref{density}), it is seen that there is a central singularity at $r=0$ for $v>0$.
The point of $v=r=0$ is more subtle but will be shown to be singular as well below.
We will study the nature of the singularity.

We have 
\begin{eqnarray}
\frac{dr}{dv}=\frac{f}{2} \label{null-eq}
\end{eqnarray}  
along a future-directed outgoing radial null geodesic, so that the region with $f<0$ is the trapped region.
A curve $f=0$ represents the trapping horizon~\cite{Hayward1994}, i.e., the orbit of the apparent horizon, which is
\begin{eqnarray}
m(v)=m_0 v^{n-3}={\tilde \alpha} r^{n-5}+r^{n-3}. \label{ah}
\end{eqnarray}  
Along the trapping horizon,
\begin{eqnarray}
ds^2=\frac{2(n-3)m_0v^{n-4}}{{\tilde \alpha}(n-5)r^{n-6}+(n-3)r^{n-4}}dv^2 
\end{eqnarray}  
is satisfied and hence it is spacelike for $v>0$ and $r>0$, i.e., it is a future outer trapping horizon, which is a local definition of black hole~\cite{Hayward1994}.  
From Eq.~(\ref{ah}), only the point $v=r=0$ may be a naked singularity in the case of $n \ge 6$, while the central singularity may be naked for $0 \le v \le v_{\rm AH}$ in the case of $n=5$, where $v_{\rm AH}$ is defined by 
\begin{eqnarray}
m_0 v_{\rm AH}^{2}={\tilde \alpha}.
\end{eqnarray}

In order to determine whether or not the singularity is naked, we investigate the future-directed outgoing geodesics emanating from the singularity. 
It is shown that if a future-directed radial null geodesic does not emanate from the singularity, then a future-directed causal (excluding radial null) geodesic does not also. (See Appendix A for the proof.)
So we consider here only the future-directed outgoing radial null geodesics.

In order to show the existence of a null geodesic emanating from the singularity, we adopt the fixed-point method~\cite{christodoulou1984,newman1986,khi2000}. 
We introduce a new coordinate $\vartheta \equiv r/(v-v_0)$ and then the null geodesic equation (\ref{null-eq}) becomes
\begin{eqnarray}
\frac{d\vartheta}{dv}+\frac{1}{(v-v_0)}(\vartheta-\eta)=\eta\Psi(v,\vartheta),
\label{theta}
\end{eqnarray}
where we have introduced a parameter $0<\eta<\infty$ and $\Psi$ is given by
\begin{eqnarray}
\Psi(v,\vartheta)&\equiv&-\frac{1}{v-v_0}+\frac{1}{2\eta (v-v_0)}\biggl[1+\frac{(v-v_0)^2\vartheta^2}{2{\tilde\alpha}} \nonumber \\
&&\times \left(1-\sqrt{1+\frac{4{\tilde\alpha}m_0v^{n-3}}{(v-v_0)^{n-1}\vartheta^{n-1}}}\right)\biggl].
\end{eqnarray}
The constant $v_0$ is zero for $n \ge 6$, while it satisfies $0 \le v_0 \le v_{\rm AH}$ for $n=5$.
If we choose the parameter $\eta$ to be $\eta=\eta_0 \equiv (1-\sqrt{m_0v_0^2/{\tilde \alpha}})/2$ for $n \ge 5$, $\Psi$ is at least $C^1$ in $v \ge v_0, \vartheta>0$.
Then we can apply the contraction mapping principle to Eq.~(\ref{theta}) to find that there exists the solution satisfying $\vartheta(v_0)=\eta_0$, and moreover that it is the unique solution of Eq.~(\ref{theta}) which is continuous at $v=v_0$. 
(See~\cite{christodoulou1984,newman1986,khi2000} for the proof.)

Now we have shown that there exists a future-directed outgoing radial null geodesic which behaves as
\begin{eqnarray}
v \simeq 2r
\label{nullray6}
\end{eqnarray}
near $v=r=0$ for $n \ge 6$, while as
\begin{eqnarray}
v \simeq v_0+\frac{2}{1-\sqrt{m_0v_0^2/{\tilde \alpha}}}r
\label{nullray51}
\end{eqnarray}
near $r=0$ and $v=v_0$ with $0 \le v_0 <v_{\rm AH}$ for $n=5$.
Along the null geodesic emanating from the singularity at $v=r=0$ for $n \ge 5$, the Kretschmann invariant $K\equiv R_{\mu\nu\rho\sigma}R^{\mu\nu\rho\sigma}$, which is represented by $f$ as 
\begin{equation}
K=(f'')^2+\frac{2(n-2)}{r^2}(f')^2+\frac{2(n-2)(n-3)}{r^4}(1-f)^2,
\label{kretchemann_e}
\end{equation}  
diverges for $r \to 0$ as 
\begin{eqnarray}
K=O(1/r^2). \label{divGB1}
\end{eqnarray}  
Along the null geodesic emanating from the singularity at $r=0$ and $0<v<v_{\rm AH}$ for $n=5$, $K$ diverges as
\begin{eqnarray}
K=O(1/r^4), \label{divGB2}
\end{eqnarray}  
and hence they are actually singular null geodesics.
We have now shown that at least a locally naked singularity is formed, and consequently the strong version of CCH is violated. 

Next we consider the structure of the naked singularity.
First we consider the case of $n \ge 6$.
We expand the trapping horizon (\ref{ah}) around $r=0$ as
\begin{eqnarray}
v&=&m_0^{-1/(n-3)}({\tilde\alpha}r^{n-5}+r^{n-3})^{1/(n-3)}, \\
&\simeq&({\tilde\alpha}/m_0)^{1/(n-3)}r^{1-2/(n-3)}.
\end{eqnarray}  
Thus, there exists a spacetime region ${\cal U}$ which is both the past of the trapping horizon and the future of the future-directed outgoing radial null geodesic $\gamma$ which behaves as Eq.~(\ref{nullray6}) near $v=r=0$.
Because the trapping horizon is spacelike for $v>0$ and $r>0$, the past-directed ingoing radial null geodesic $\zeta$ emanating from an event in ${\cal U}$ never crosses the trapping horizon.
Also $\zeta$ never crosses $\gamma$ except for $v=r=0$ because they are both future-directed outgoing radial null geodesics.
Consequently, $\zeta$ inevitably reaches the singularity at $v=r=0$.
Since ${\cal U}$ is an open set, it is concluded that there exists an infinite number of future-directed outgoing radial null geodesics emanating from the singularity at $v=r=0$.
Such geodesics should correspond to the solution of Eq.~(\ref{theta}) with $\eta=0$ or $\eta=\infty$.
On the other hand, the future-directed ingoing radial null geodesic terminating at $v=r=0$ is only $v=0$, thus we conclude that the naked singularity at $v=r=0$ has the ingoing-null structure.

In the case of $n=5$, the singularity at $r=0$ and $v=v_0$ with $0 \le v_0 <v_{\rm AH}$ is both an endpoint of the future-directed ingoing radial null geodesic and the initial point of the future-directed outgoing radial null geodesic, therefore we conclude that the naked singularity has the timelike structure.

On the other hand, it cannot be concluded from the present analysis that there exist no null geodesics emanating from the singularity at $r=0$ and $v=v_{\rm AH}$ because null geodesics which correspond to the solution of Eq.~(\ref{theta}) with $\eta=0$ or $\eta=\infty$ might exist. 
More detailed analyses are needed in order to clarify the nature of that singularity at $r=0$ and $v=v_{\rm AH}$.
If there is a null geodesic emanating from the point $r=0$ and $v=v_{\rm AH}$, it is concluded by the same discussion as that in the case of $n \ge 6$ that the singularity has an ingoing null portion.

We can consider the situation in which the null dust fluid is turned off at a finite time $v=v_{\rm f}>0$, whereupon the field settles into the GB-Schwarzschild spacetime with $m=m_0 v_{\rm f}^{n-3}$, which is asymptotically flat.
The Minkowski spacetime for $v<0$ is joined to the GB-Schwarzschild spacetime for $v>v_{\rm f}$ by way of the GB-Vaidya solution (\ref{f-GB}) (Fig.~\ref{Fig1}).
Then the singularity is always globally naked when $n=5$ with $v_{\rm f} \le v_{\rm AH}$ since no horizon is formed.
When $n \ge 6$ or $n=5$ with $v_{\rm f} > v_{\rm AH}$, the singularity can also be globally naked.
If we set $v_{\rm f}$ to a sufficiently small value and a sufficiently close value to $v_{\rm AH}$ for $n \ge 6$ and $n=5$, respectively, a null ray emanating from the singularity reaches the surface $v=v_{\rm f}$ in the untrapped region, and consequently it can escape to infinity.
In that case, the singularity is globally naked, i.e., the weak version of CCH is violated.
In order to determine the values of the parameters with which the singularity is globally naked, numerical integrations of the null geodesic equation (\ref{null-eq}) are required, in which the asymptotic solutions (\ref{nullray6}) and (\ref{nullray51}) are used as the initial conditions.

\begin{figure}[tbp]
\includegraphics[width=.99\linewidth]{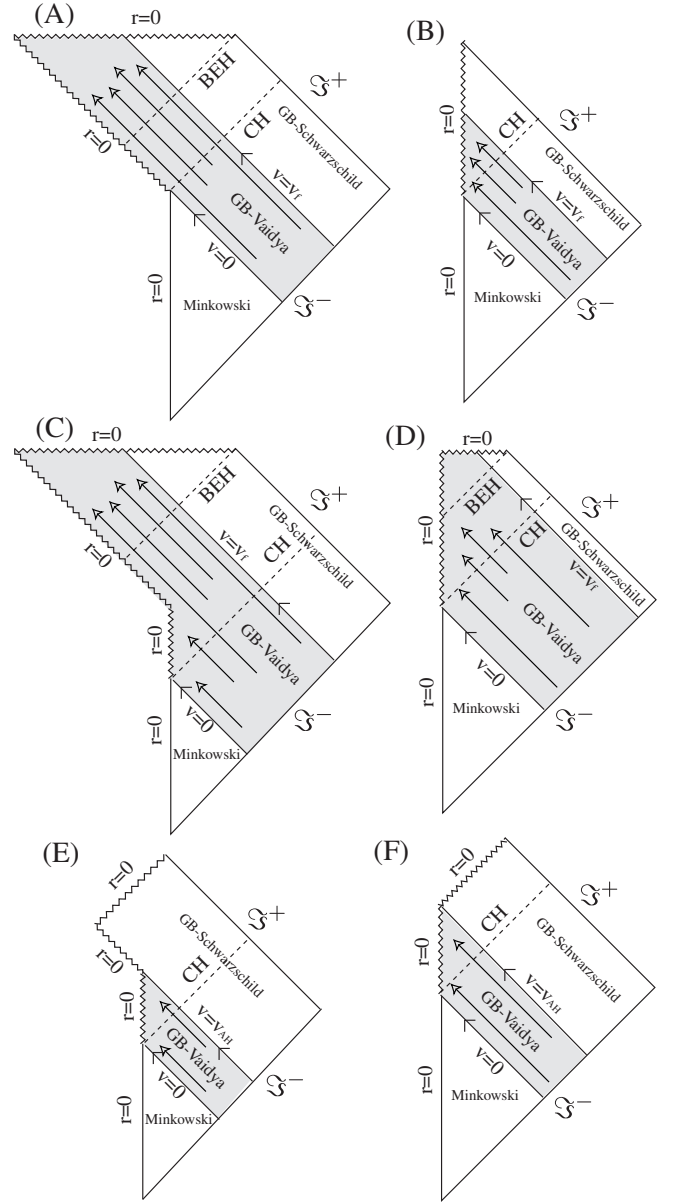}
\caption{
Possible global structures of the globally naked singularity formation.
Zigzag lines represent the central singularities. 
$\Im^{+(-)}$ corresponds to the future (past) null infinity. 
BEH and CH stand for the black-hole event horizon and the Cauchy horizon, respectively.
(See Ref.~\cite{he1973} for rigorous definitions of the event horizon and the Cauchy horizon.)
Naked singularity formation in general relativity is represented by (A).
In 5-dimension with the Gauss-Bonnet term, the global structure is represented by either (C) or (D) ((E) or (F)) for $v_{\rm f} > (=) v_{\rm AH}$, while it is represented by (B) for $v_{\rm f} < v_{\rm AH}$.
In $n(\ge 6)$-dimension with the Gauss-Bonnet term, the global structure is represented by (A). 
There are counterparts of (A), (C), and (D) which represent the locally naked singularity formation.
}
\label{Fig2}
\end{figure}

The structures of the singularity in the outer GB-Schwarzschild spacetime are as follows~\cite{tm2005}.
The singularity is spacelike for $n \ge 6$ and for $n=5$ with $v_{\rm f}>v_{\rm AH}$.
It is timelike for $n=5$ with $v_{\rm f}<v_{\rm AH}$, while it is null for $n=5$ with $v_{\rm f}=v_{\rm AH}$.
Together with these results, the possible global structures of the gravitational collapse are drawn in Fig.~\ref{Fig2}.

Next we consider the general relativistic case, i.e., the $n$-dimensional Vaidya solution.
In this case, the trapping horizon is represented by $v=v_{{\rm h}}r$, where 
\begin{eqnarray}
v_{{\rm h}} \equiv \frac{1}{m_0^{1/(n-3)}}, \label{v1ah}
\end{eqnarray}
and hence only the point $v=r=0$ may be naked.
We find the exact power-law solution of the null geodesic equation (\ref{null-eq}):
\begin{eqnarray}
v = v_1 r, \label{nullraygr}
\end{eqnarray}
when 
\begin{eqnarray}
g(v_1) \equiv m_0v_1^{n-2}-v_1+2=0 \label{geq}
\end{eqnarray}
has at least one positive real root.
This condition is realized when 
\begin{eqnarray}
g(v_{{\rm ex}}) \le 0 \label{cond2}
\end{eqnarray}
holds, where $v_{{\rm ex}} \equiv 1/[(n-2)m_0]^{1/(n-3)}$ satisfies
\begin{eqnarray}
\frac{dg(v_{{\rm ex}})}{dv_1}=0.
\end{eqnarray}
The condition (\ref{cond2}) is written as
\begin{eqnarray}
m_0 \le \frac{(n-3)^{n-3}}{2^{n-3}(n-2)^{n-2}}. \label{condgr}
\end{eqnarray}
It is emphasized that Eq.~(\ref{nullraygr}) is not an asymptotic solution but an exact solution of the null geodesic equation (\ref{null-eq}).
Along this null geodesic, the Kretschmann invariant diverges for $r \to 0$ as 
\begin{eqnarray}
K=O(1/r^4), \label{divGR1}
\end{eqnarray}  
and hence it is concluded that an ingoing-null naked singularity is formed at $v=r=0$.
This result is consistent with the results in~\cite{gs2000,gd2001}.

Here we show that if Eq.~(\ref{condgr}) is satisfied then the null geodesic (\ref{nullraygr}) can reach infinity and therefore the singularity is globally naked independent of the value of $v_{\rm f}$.
Because of
\begin{eqnarray}
\frac{dg}{dv_1}=(n-2)m_0v_1^{n-3}-1
\end{eqnarray}
and $g(0)=2>0$, the null geodesic equation (\ref{null-eq}) represents either one or two solutions when Eq.~(\ref{condgr}) is satisfied.
We denote them as $v=v_{a}r$ and $v=v_{b}r$ with $0<v_{a} \le v_{{\rm ex}} \le v_{b}$, where we have equality when $g(v_{{\rm ex}})=0$.
From Eq.~(\ref{v1ah}), $v_{{\rm h}}>v_{{\rm ex}} \ge v_{a}$ satisfies, so that the null geodesic $v=v_{a} r$ is always in the untrapped region and reaches the future null infinity independent of the value of $v_{\rm f}$.
Thus, when Eq.~(\ref{condgr}) is satisfied, the solution represents the formation of a globally naked singularity, i.e., the weak CCH is violated. 

In the case that $m_0$ does not satisfy the condition (\ref{condgr}), more detailed analyses are needed to determine whether the singularity is naked or censored because it might be possible that there exist null geodesics which do not obey the power law near $v=r=0$.

\section{Curvature strength of a naked singularity}
In the previous section, it was shown that a naked singularity is inevitably formed in Gauss-Bonnet gravity.
In this section, we investigate the curvature strength of the naked singularity.
We define 
\begin{equation}
\psi \equiv R_{\mu\nu}k^\mu k^\nu, \label{psi}
\end{equation}
where $k^\mu \equiv dx^\mu/d\lambda$ is the tangent vector of the future-directed outgoing radial null geodesic which emanates from the singularity and is parameterized by an affine parameter $\lambda$.
We evaluate the strength of the naked singularity by the dependence of $\psi$ on $\lambda$ near the singularity.

In the 4-dimensional case, the strong curvature condition (SCC)~\cite{tipler1977} and the limiting focusing condition (LFC)~\cite{krolak1987} are defined.
Following the work of Clarke and Kr\'{o}lak~\cite{ck1985}, consider a geodesic ($\gamma$) with tangent vector $k^{\mu}$ and terminating at or emanating from a singularity, where $\lambda=0$. 
If $\lim_{\lambda\to 0}\lambda^{2}\psi >0$ and $\lim_{\lambda\to 0}\lambda \psi >0$, then SCC and LFC are satisfied along $\gamma$, respectively.
However, these results have not been extended to the higher-dimensional case so far.

From Eq.~(\ref{null-eq}), a relation 
\begin{equation}
k^r=\frac{f}{2}k^v
\end{equation}
is obtained.
Using this, we obtain from Eq.~(\ref{psi}) that 
\begin{equation}
\psi=-\frac{2(n-2){\dot f}}{rf^2}(k^r)^2. \label{beq100}
\end{equation}
We write the geodesic equation for the radial null geodesic:
\begin{eqnarray}
\frac{d}{d\lambda}k^r=\frac{2{\dot f}}{f^2}(k^r)^2.\label{neq5}
\end{eqnarray}

First, we consider the case in Gauss-Bonnet gravity.
Along the null geodesics (\ref{nullray6}) and (\ref{nullray51}), we obtain
\begin{eqnarray}
\lim_{r \to 0}f&=&1, \label{limit1} \\
\lim_{r \to 0}{\dot f}&=&-\frac{n-3}{2}\sqrt{\frac{2^{n-5}m_0}{{\tilde \alpha}}} \label{limit2}
\end{eqnarray}
for $n \ge 6$ and
\begin{eqnarray}
\lim_{r \to 0}f&=&1-\sqrt{\frac{m_0v_0^2}{{\tilde\alpha}}}, \label{limit3} \\
\lim_{r \to 0}{\dot f}&=&-\sqrt{\frac{m_0}{{\tilde \alpha}}} \label{limit4}
\end{eqnarray}
for $n=5$, respectively.
Using Eqs.~(\ref{limit1})--(\ref{limit4}), we find that Eq.~(\ref{neq5}) is reduced to the following form near the singularity:
\begin{eqnarray}
\frac{d}{d\lambda}k^r \simeq -q(k^r)^2, \label{neq6}
\end{eqnarray}
where $q$ is defined by
\begin{eqnarray}
q \equiv (n-3)\sqrt{\frac{2^{n-5}m_0}{{\tilde \alpha}}}
\end{eqnarray}
for $n \ge 6$, while by
\begin{eqnarray}
q \equiv \frac{2\sqrt{m_0/{\tilde\alpha}}}{(1-\sqrt{m_0v_0^2/{\tilde\alpha}})^2}
\end{eqnarray}
for $n=5$.
The solution of Eq.~(\ref{neq6}) is
\begin{eqnarray}
r \simeq \frac{1}{q}\ln(C\lambda+D),\label{nullGB6}
\end{eqnarray}
where $C$ and $D$ are integration constants. Here, we set $C=D=1$ without loss of generality by redefinition of the affine parameter so that $\lambda=0$ corresponds to the singularity $r=0$.

After straightforward calculations using Eqs.~(\ref{limit1})--(\ref{limit4}) and (\ref{nullGB6}), we obtain from Eq.~(\ref{beq100}) that
\begin{eqnarray}
\lim_{\lambda \to 0}\psi \simeq \frac{n-2}{\ln(\lambda+1)} \label{strengthGB6}
\end{eqnarray}
for $n \ge 5$. 
Thus, 
\begin{eqnarray}
\lim_{\lambda \to 0}\lambda \psi = n-2
\end{eqnarray}
is satisfied for $n \ge 5$.

Next, we consider the case in general relativity.
In this case, we obtain from Eq.~(\ref{neq5}) that the null geodesic (\ref{nullraygr}) obeys the null geodesic equation
\begin{eqnarray}
\frac{d}{d\lambda}k^r = -\frac{s}{r}(k^r)^2,\label{beqgr}
\end{eqnarray}
where 
\begin{eqnarray}
s \equiv \frac{(n-3)m_0v_1^{n-2}}{2}.
\end{eqnarray}
The solution of Eq.~(\ref{beqgr}) is
\begin{eqnarray}
r=(E\lambda+F)^{1/(1+s)},
\end{eqnarray}
where $E$ and $F$ are integration constants. We set $E=1$ and $F=0$ by redefinition of the affine parameter.
After calculations in a manner similar to those in Gauss-Bonnet gravity, we obtain
\begin{equation}
\psi = \frac{(n-2)s}{(1+s)^2\lambda^2}. \label{strengthGR}
\end{equation}
This result is consistent with the results in~\cite{gs2000,gd2001}.

Consequently, it is concluded by comparing Eq.~(\ref{strengthGB6}) with Eq.~(\ref{strengthGR}) that the strength of the naked singularity in Gauss-Bonnet gravity is weaker than that in general relativity.

\section{Discussion and conclusions}
We have obtained an exact solution in Gauss-Bonnet gravity, which represents the spherically symmetric gravitational collapse of a null dust fluid in $n(\ge 4)$-dimensions.
For $n \ge 5$, the solution is reduced to the $n$-dimensional Vaidya-(anti)de Sitter solution in the general relativistic limit. 
For $n=4$, the Gauss-Bonnet term does not contribute to the field equations, so that the solution coincides with the Vaidya-(anti)de Sitter solution.
Applying the solution to the situation in which a null dust fluid radially injects into an initially Minkowski region, we have investigated the effects of the Gauss-Bonnet term on the final fate of the gravitational collapse.

We have assumed that the mass function has the form of $m(v)=m_0v^{n-3}$.
Then, it has been found that there always exists a future-directed outgoing radial null geodesic emanating from the singularity in Gauss-Bonnet gravity, i.e., a naked singularity is inevitably formed.
On the other hand, in the general relativistic case, there exists such a null geodesic only when $m_0$ takes a sufficiently small value.
This result implies that the effects of the Gauss-Bonnet term on gravity worsen the situation from the viewpoint of CCH rather than prevent naked singularity formation.

Furthermore, the Gauss-Bonnet term drastically changes the nature of the singularity and the whole picture of gravitational collapse.
The picture of the gravitational collapse for $n=5$ is quite different from that for $n\ge 6$.
For $n\ge 6$, as well as the general relativistic case for $n \ge 4$, a massless ingoing null naked singularity is formed.
On the other hand, for the special case $n=5$, a massive timelike naked singularity is formed. 
It is found from Eq.~(\ref{ah}) that the formation of a massive timelike singularity in 5-dimensions is generic for the general mass function $m(v)$ which satisfies $m(0)=0$ and ${\dot m} \ge 0$.

Here, we should mention the substantial work by Lake in 4-dimensional spherically symmetric spacetimes~\cite{lake1992}.
It has been shown under very generic situations that the massive singularities formed from regular initial data are censored~\cite{lake1992}.
In other words, if a naked singularity is formed from regular initial data, it is massless or with negative mass.
This result was obtained without using the Einstein equations and may be extended into higher dimensions.
One might think, therefore, that this result is inconsistent with ours in Gauss-Bonnet gravity with $n=5$.
However, the discussion and results in~\cite{lake1992} cannot apply directly to our system.
In Lake's analysis, the Misner-Sharp mass is adopted to evaluate the mass of the singularity.
On the other hand, we adopted $m(v)$, which is different from the higher-dimensional Misner-Sharp mass.
As shown in Appendix~B, $m(v)$ is preferable in Gauss-Bonnet gravity to evaluate the mass of the singularity.

Although naked singularities are inevitably formed in Gauss-Bonnet gravity, the Gauss-Bonnet term makes the strength of the naked singularity weaker than that in the general relativistic case.
In association with this, there does exist a possible formulation of CCH, which asserts that the formation of weak naked singularities need not be ruled out~\cite{tce1980}.
In this sense, the Gauss-Bonnet term works well in the spirit of CCH.

Lastly, several notes should be made.
In this paper, we have assumed $m(v)=m_0v^{n-3}$ for simplicity.
Analyses with a more general form of $m(v)$ are needed in order to determine whether the nature of the singularities and the picture of gravitational collapse obtained in this paper are generic or not.
Also, the higher-order correction terms to gravity might change the nature of singularity and the final fate of gravitational collapse.
In addition, the assumption of spherical symmetry is very restricted.
Very recently, the naked singularity formation in the self-similar Vaidya solution was found to be stable against non-spherical linear perturbations with even parity~\cite{nw2005}.
Analyses in Gauss-Bonnet gravity beyond spherical symmetry must be addressed.
These ideas are under investigation.

\acknowledgments
I am very grateful to anonymous referees for their useful comments. 
I would like to thank T.~Torii, U.~Miyamoto, T.~Harada, and M.~Nozawa for discussion and comments. 
I would also like to thank N.~Ohta for comments on references.

~\\ {\em Note added:}

After completing this work we were informed that Kobayashi also obtained the solution (\ref{f-eq}) and its generalization to the spacetime ${\ma M} \approx {\ma M}^2 \times {\ma K}^{n-2}$, where ${\ma K}^{n-2}$ is the $(n-2)$-dimensional Einstein space~\cite{kobayashi2005}. 

~\\

\appendix

\section{geodesics from singularities}
\label{geodesics}
In this appendix, we show that {\it if a future-directed causal (excluding radial null) geodesic emanates from the central singularity, then a future-directed radial null geodesic emanates from the central singularity}.
The contraposition of this theorem implies that it is sufficient to consider only the future-directed outgoing radial null geodesics in order to determine whether or not the singularity is naked.
The proof is similar to that in the four-dimensional case in~\cite{nmg2002}.

In the GB-Vaidya spacetime 
\begin{equation}
ds^2=-fdv^2+2dvdr+r^2d\Omega_{n-2}^2,
\end{equation}  
where $f$ is given by Eq.~(\ref{f-GB}), the tangent to a causal geodesic satisfies
\begin{eqnarray}
-f\left(\frac{dv}{d\lambda}\right)^2+2\frac{dv}{d\lambda}\frac{dr}{d\lambda}+\frac{L^2}{r^2}=\epsilon, \label{hamilton}
\end{eqnarray}  
where $\lambda$ is an affine parameter, $L$ is the conserved angular momentum, and $\epsilon=0,-1$ for null and timelike geodesics, respectively.
Then, at any point on such a geodesic,
\begin{eqnarray}
f\left(\frac{dv}{d\lambda}\right)^2 \ge 2\frac{dv}{d\lambda}\frac{dr}{d\lambda}
\end{eqnarray}
with equality holding only for radial null geodesics.
For the future-directed outgoing geodesics, this gives
\begin{eqnarray}
\frac{dv}{dr}\ge \frac{2}{f}>0
\end{eqnarray}
and 
\begin{eqnarray}
\frac{dv}{dr}\le \frac{2}{f}<0
\end{eqnarray}
in the untrapped and trapped regions, respectively.
Therefore, 
\begin{eqnarray}
\frac{dv_{\rm CG}}{dr}>\frac{dv_{\rm RNG}}{dr}>0 \label{geodesics1}
\end{eqnarray}
and 
\begin{eqnarray}
\frac{dv_{\rm CG}}{dr}<\frac{dv_{\rm RNG}}{dr}<0 \label{geodesics2}
\end{eqnarray}
are satisfied in the untrapped and trapped regions, respectively, where the subscripts represent causal (excluding radial null) geodesics and outgoing radial null geodesics, respectively.

Let us consider the $(r,v)$-plane.
The singularity is located at $r=0$ for $v \ge 0$.
First we consider the singularity with $f<0$.
By Eq.~(\ref{geodesics2}), the past-directed ingoing geodesics emanating from an event with $r>0$ in the trapped region cannot reach the singularity at $r=0$.
Therefore, there is no future-directed outgoing geodesic emanating from the singularity in the trapped region, i.e., the singularity with $f<0$ is censored.

Next we focus on the singularity with $0\le v \le v_{\rm AH}$, where $f\ge 0$ holds.
Now suppose that $v=v_{\rm CG}(r)$ extends back to a central singularity located at $(r,v)=(0,v_{s})$, where $v_{s}$ satisfies $0 \le v_{s} \le v_{\rm AH}$.
There exists a portion of $v=v_{\rm CG}(r)$ which is in the untrapped region.
Let $p$ be any point on such a potion of $v=v_{\rm CG}(r)$ and to the future of the singularity.
Applying inequality (\ref{geodesics1}) at $p$, we see that $v=v_{\rm RNG}(r)$ through $p$ crosses $v=v_{\rm CG}(r)$ from above and hence the points on $v=v_{\rm RNG}(r)$ prior to $p$ must lie to the future of points on $v=v_{\rm CG}(r)$ prior to $p$, in the sense $v_{\rm RNG}(r)>v_{\rm CG}(r)$ for $r \in (0,r_{\ast})$, where $r_{\ast}$ corresponds to $p$.
Thus, the radial null geodesics must extend back to $r=0$ with $v=v_0$ satisfying $v_s \le v_{0} \le v_{\rm AH}$, and so must emerge from the singularity.

It is noted that $v_{\rm AH}=v_s=v_{0}=0$ holds in the case of $n \ge 6$.

\section{mass of naked singularity}
\label{qlm}
In this appendix, we show that the higher-dimensional Misner-Sharp mass is not an appropriate quasi-local mass in Gauss-Bonnet gravity.

Firstly, we give a definition of the higher-dimensional Misner-Sharp mass.
We consider the $n$-dimensional spherically symmetric spacetime ${\ma M} \approx {\ma M}^2\times {\ma S}^{n-2}$ with the general metric 
\begin{eqnarray}
g_{\mu\nu}=\mbox{diag}(g_{AB},r^2\gamma_{ij}),
\label{eq:structure}
\end{eqnarray}
where $g_{AB}$ is an arbitrary Lorentz metric on ${\ma M}^2$, $r$ is a scalar function on ${\ma M}^2$ with $r=0$ defining the boundary of ${\ma M}^2$, and $\gamma_{ij}$ is the unit curvature metric on ${\ma S}^{n-2}$. 
The higher-dimensional Misner-Sharp mass $m_{\rm MS}$ is a scalar on ${\ma M}^2$ defined by
\begin{eqnarray}
\label{qlm}
m_{\rm MS} \equiv \frac{(n-2)V_{n-2}}{2\kappa_n^2}r^{n-3}(1-r_{,A}r^{,A}).
\end{eqnarray}  

There are several properties which a well-defined quasi-local mass should satisfy~\cite{hayward1996,szabados2004}. 
One of them states that there is an invariant mass function in spherically symmetric spacetime, that any definition of the quasi-local mass should be reduced to in the special case with spherical symmetry.
In particular, in the Schwarzschild spacetime of mass $M_{\rm Sch}$, the invariant mass function should coincide with $M_{\rm Sch}$ because it is the unique vacuum solution in general relativity.
In Gauss-Bonnet gravity, on the other hand, the unique spherically symmetric vacuum solution is the GB-Schwarzschild solution.
Therefore, a well-defined quasi-local mass in Gauss-Bonnet gravity should be reduced to the mass for this solution in the spherically symmetric vacuum case.

The metric of the GB-Schwarzschild solution is given by
\begin{equation}
ds^2=-fdv^2+2dvdr+r^2d\Omega_{n-2}^2
\end{equation}  
with
\begin{equation}
f=1+\frac{r^2}{2\tilde{\alpha}}\Biggl(1-\sqrt{1+4\tilde{\alpha}\frac{m}{r^{n-1}}}\Biggr), 
\end{equation}  
where $m$ is a constant.
The $n$-dimensional Schwarzschild solution is obtained in the limit of $\tilde{\alpha} \to 0$ as
\begin{equation}
f=1-\frac{m}{r^{n-3}},  \label{f-GR}
\end{equation}  
of which mass is given by $M_{\rm Sch} \equiv (n-2)V_{n-2}m/(2\kappa_n^2)$, where $V_{n-2}$ is the volume of an $(n-2)$-dimensional unit sphere.
On the other hand, the mass for the GB-Schwarzschild solution also coincides with $M_{\rm Sch}$~\cite{massGB}.
Thus, a well-defined quasi-local mass should be reduced to $M_{\rm Sch}$ for the spherically symmetric vacuum case in Gauss-Bonnet gravity as well as in general relativity.

From Eq.~(\ref{qlm}), we can show $m_{\rm MS}=M_{\rm Sch}$ for the $n$-dimensional Schwarzschild solution, while we obtain
\begin{equation}
\label{qlm1}
m_{\rm MS} = -\frac{(n-2)V_{n-2}}{4\tilde{\alpha}\kappa_n^2}r^{n-1}\Biggl(1-\sqrt{1+4\tilde{\alpha}\frac{m}{r^{n-1}}}\Biggr)
\end{equation}
for the GB-Schwarzschild solution.
Indeed, $m_{\rm MS}$ does not coincide with $M_{\rm Sch}$ for $n \ge 5$ for the latter case.
Consequently, $m_{\rm MS}$ is not an appropriate quasi-local mass in Gauss-Bonnet gravity.

In this paper, $m(v)$ in Eq.~(\ref{f-GB}) is adopted in order to evaluate the mass of the singularity.
$m(v)$ is preferable rather than $m_{\rm MS}$ in Gauss-Bonnet gravity because it is reduced to $2\kappa_n^2M_{\rm Sch}/[(n-2)V_{n-2}]$ for the static limit, i.e., in the GB-Schwarzschild solution.


\end{document}